\documentclass[prl,aps,twocolumn,superscriptaddress,showpacs]{revtex4}
\usepackage{amsmath,amssymb,graphicx}
\usepackage{pxfonts}
\usepackage{graphicx}
\usepackage{color}
\usepackage{float}
\begin{document}

\date{\today}

\title{High-mobility indirect excitons in wide single quantum well}

\author{C.~J.~Dorow} \email{cdorow@physics.ucsd.edu} \author{M.~W.~Hasling} \author{D.~J.~Choksy} \author{J.~R.~Leonard} \author{L.~V.~Butov} 
\affiliation{Department of Physics, University of California at San Diego, La Jolla, CA 92093, USA}
\author{K.~W.~West} \author{L.~N.~Pfeiffer} \affiliation{Department of Electrical Engineering, Princeton University, Princeton, New Jersey 08544, USA}

\begin{abstract}
Indirect excitons (IXs) are bound pairs of electrons and holes confined in spatially separated layers. We present wide single quantum well (WSQW) heterostructures with high IX mobility, spectrally narrow IX emission, voltage-controllable IX energy, and long and voltage-controllable IX lifetime. This set of properties shows that WSQW heterostructures provide an advanced platform both for studying basic properties of IXs in low-disorder environments and for the development of high-mobility excitonic devices.
\end{abstract}

\maketitle

Spatially indirect excitons (IXs) are formed by electrons and holes confined in separated layers. The spatial separation between the electron and hole layers allows one to control the overlap of electron and hole wave functions and achieve long IX lifetimes that are orders of magnitude longer than those of direct excitons~\cite{Lozovik1976}. Due to their long lifetimes, IXs can cool below the temperature of quantum degeneracy~\cite{Butov2001}. The realization of cold IXs in coupled quantum well (CQW) heterostructures led to the observation of spontaneous coherence and condensation of IXs~\cite{High2012} and perfect Coulomb drag~\cite{Nandi2012}. A set of phenomena was found in the IX condensate, including the spatially modulated exciton state~\cite{Butov2002, Alloing2014}, commensurability effect of exciton density wave~\cite{Yang2015}, spin textures~\cite{High2013}, and Pancharatnam-Berry phase and long-range coherent spin transport~\cite{Leonard2018}.

The long IX lifetimes also allow IXs to travel over large distances before recombination~\cite{Hagn1995, Butov1998, Larionov2000, Butov2002, Voros2005, Ivanov2006, Gartner2006, Hammack2009, Leonard2009, Lasic2010, Alloing2012, Lasic2014, Finkelstein2017}. A set of exciton transport phenomena was observed, including the inner ring in exciton emission patterns~\cite{Butov2002, Ivanov2006, Hammack2009, Alloing2012}, exciton localization-delocalization transition in random~\cite{Butov2002, Ivanov2006, Hammack2009}, periodic~\cite{Remeika2012, Remeika2015}, and moving~\cite{Winbow2011} potentials, coherent exciton transport with suppressed scattering~\cite{High2012}, and transistor effect for excitons~\cite{High2008, Andreakou2014}.

IXs have a built-in dipole moment $ed$, where $d$ is the separation between the electron and hole layers. Gate voltage $V_{\rm g}$ changes the IX energy by $edF_z$ ($F_z \propto V_{\rm g}$ is electric field  perpendicular to the QW plane created by voltage)~\cite{Miller1985, Polland1985}. This allows creating tailored in-plane potential landscapes for IXs $E(x,y) = - edF_z(x,y)$ and controlling them in situ by voltage $V_{\rm g}(x,y)$. A variety of electrostatic potential landscapes, including traps~\cite{Huber1998, Gorbunov2004, Hammack2006, Chen2006, High2009nl, High2012nl, Schinner2013, Shilo2013, Mazuz-Harpaz2017}, static~\cite{Remeika2012, Remeika2015, Zimmermann1997, Zimmermann1998} and moving~\cite{Winbow2011} lattices, ramps~\cite{Hagn1995, Gartner2006, Dorow2016}, narrow channels~\cite{Vogele2009, Cohen2011}, and split gate devices~\cite{Dorow2018}, was created for studying basic exciton properties. 

IX devices are also explored for applications in signal processing based on novel computational state variables beyond magnetism or charge. Excitonic devices possess potential advantages over electronic devices: (i) Excitons are bosons and can form a coherent condensate with vanishing resistance for exciton currents and low switching voltage for excitonic transistors due to suppressed thermal tails. This gives the opportunity to develop energy-efficient computation based on excitons. (ii) Excitons can directly transform to photons providing the possibility for efficient coupling of excitonic signal processing to optical communication. (iii) The sizes of excitonic devices scale by the exciton radius and de Broglie wavelength, which are much smaller than the photon wavelength, so excitonic circuits may be created at sub-photon-wavelength scales. Experimental proof-of-principle demonstrations have been performed for excitonic ramps (excitonic diodes)~\cite{Hagn1995, Gartner2006, Dorow2016}, excitonic conveyers (excitonic CCD)~\cite{Winbow2011}, and excitonic transistors~\cite{High2008, Andreakou2014}.

The realization of the excitonic devices relies on meeting the following requirements: (1) IX energy is controlled by voltage, (2) IX recombination rate is controlled by voltage and long IX lifetimes are achieved, (3) long-range IX transport over lengths exceeding the in-plane dimensions of excitonic devices is achieved. All these requirement were met with IXs in GaAs CQWs where electrons were confined in one QW and holes in the other QW and the studies outlined above used the CQW platform. 

The crucial issue both for studying basic properties of IXs and for the development of excitonic devices is in-plane disorder. Since the energy of a particle in a QW scales with the QW width $L$ roughly as $1/L^2$ random QW width fluctuations generally cause a smaller disorder in a wider QW. Therefore, wide QWs may offer advantages for creating low-disorder IX devices. Although GaAs wide single quantum wells (WSQWs) were probed since the pioneering studies of IXs~\cite{Miller1985, Polland1985}, no long-range IX transport was demonstrated in the WSQW devices. In this paper, we report on the studies of WSQW heterostructures which meet all the above requirements and demonstrate high IX mobility, spectrally narrow IX emission, voltage-controllable IX energy, and long and voltage-controllable IX lifetime.

\begin{figure}[htbp]
\includegraphics[width=8cm]{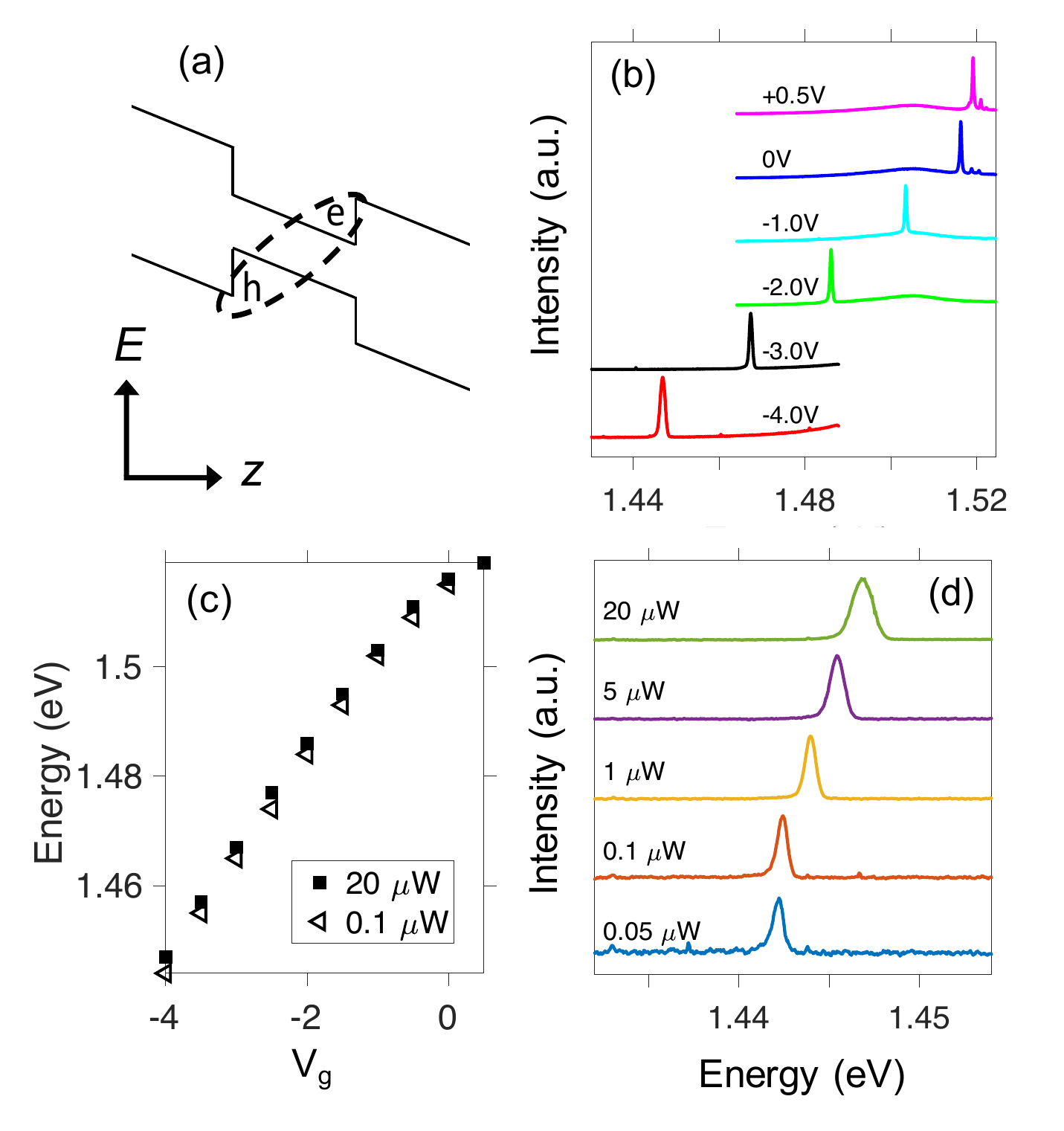}
\caption{(a) WSQW band diagram. An oval indicates an indirect exciton (IX) composed of an electron ($e$) and a hole ($h$). (b) Emission spectra vs voltage $V_{\rm g}$ for laser excitation power $P_{\rm ex} = 20$~$\mu$W. (b) IX energy vs $V_{\rm g}$ for $P_{\rm ex} = 0.1$ and 20~$\mu$W. IX spectrum vs $P_{\rm ex}$ for $V_{\rm g} = -4$~V. All spectra are spatially integrated.}
\end{figure}

The studied WSQW heterostructures are grown by molecular beam epitaxy (Fig.~1a). An $n^+$-GaAs layer with n$_{\rm Si} = 10^{18}$~cm$^{-3}$ serves as a bottom electrode. Single 35~nm GaAs QW is positioned 0.2~$\mu$m above the $n^+$-GaAs layer within an undoped 1~$\mu$m thick Al$_{0.3}$Ga$_{0.7}$As layer. The WSQW is positioned closer to the homogeneous bottom electrode to suppress the fringing in-plane electric field $F_r$ in excitonic devices~\cite{Hammack2006}. Otherwise, a high $F_r$ could lead to IX dissociation~\cite{Zimmermann1997}. The top semitransparent electrode is fabricated by applying 2~nm Ti and 7~nm Pt.

In cw experiments, excitons are generated by a 633~nm HeNe laser focused to a spot with a half width at half maximum (HWHM) $\sim 7$~$\mu$m. Photoluminescence (PL) is measured by a spectrometer and a liquid-nitrogen-cooled charge coupled device camera (CCD). 

Time-resolved optical imaging is performed using a pulsed laser excitation. IXs are generated by a 640-nm laser with a pulse duration of $\tau_{\rm width} = 2000$~ns and a pulse period of $\tau_{\rm pulse} = 6000$~ns with an edge sharpness of $\sim 1$~ns (Fig.~3a). The rectangular-shaped pulses are realized by a pulse generator driving a semiconductor laser. The pulse duration and period are optimized to allow the IX PL image to reach equilibrium during the laser excitation and decay between laser pulses. The laser is focused to a $R_0 = 5$~$\mu$m HWHM spot. Emission images are integrated over $5$-ns windows ($\delta t = 5$~ns) and taken for delay times $t$ after the onset of the laser pulse, defined such that a delay time $t$ corresponds to an image taken during time $t - \delta t$ to $t$. The PL images are captured using a PicoStar HR TauTec time-gated intensifier. The PL passes through a spectrometer with a resolution of 0.18~meV before entering the intensifier couplead to a liquid-nitgoren-cooled CCD in order to obtain spectral resolution. The spectrally and time-resolved imaging enables the direct measurement of the evolution of the IX PL as a function of delay time $t$. The measurements are performed at $T_{\rm bath} = 1.7$~K.

A narrow IX PL line is observed on a background of spectrally wide emission of $n^+$-GaAs layer (Fig.~1b). The IX energy is effectively controlled by applied voltage $V_{\rm g}$ (Fig.~1b,c). The IX energy shift with voltage in WSQW in this experiment $\sim 70$~meV is comparable to the IX energy shift in CQW heterostructures used in earlier excitonic devices~\cite{High2008}, indicating that tailored in-plane potential landscapes and, in turn, excitonic devices can be created by voltage for IXs in WSQW as efficiently as for IXs in CQW. In a wide range of $V_{\rm g}$, the IX energy shift with voltage $edF_z$ is close to linear and corresponds to the IX dipole with $d \approx 19$~nm (Fig.~1c).

The IX energy increases with the density, which is controlled by the excitation power~(Fig.~1d). This energy enhancement $\Delta E$ corresponds to the repulsive interaction between the dipolar IXs~\cite{Remeika2015}. $\Delta E$ can be used for estimating the IX density $n$. For instance, for $\Delta E = 1$~meV and $d = 19$~nm, a rough estimate for $n$ using the plate capacitor formula $\Delta E = 4\pi e^2 d n / \varepsilon$ gives $n \sim 4 \times 10^{9}$~cm$^{-2}$, this estimate can be improved taking into account IX correlations~\cite{Remeika2015}. The IX linewidth increases with increasing density due to interaction-induced broadening~\cite{High2009nl}. For spatially integrated spectra, the IX energy variation with the distance from the origin (discussed below) also contributes to the linewidth.

\begin{figure}[htbp]
\includegraphics[width=7.5cm]{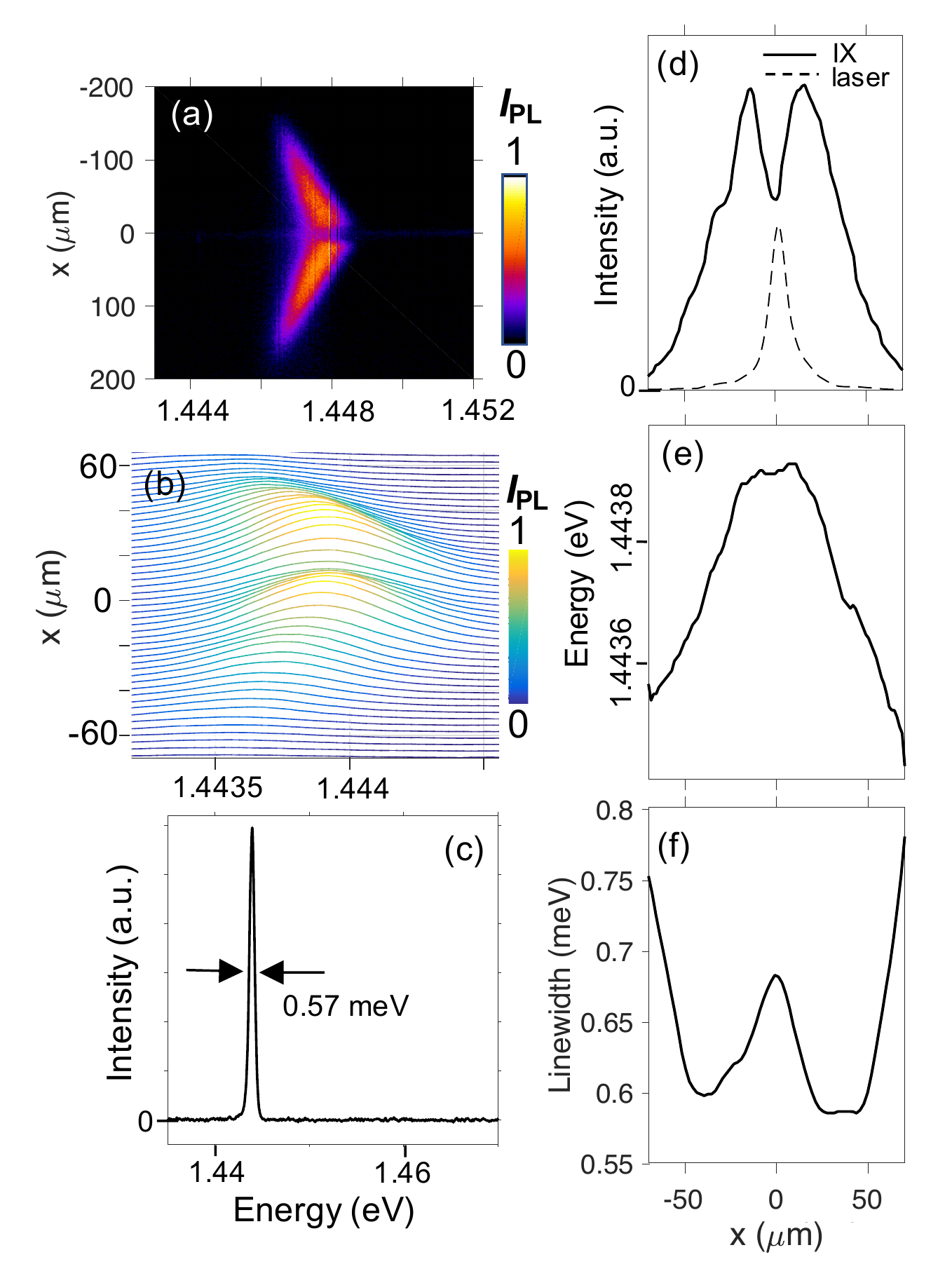}
\caption{(a) $x$-energy IX emission pattern at cw excitation. $P_{\rm ex} = 20$~$\mu$W and $V_{\rm g} = -4$~V. Laser excitation spot with HWHM = 7~$\mu$m is centered at $x = 0$. (b) The spatial dependence of IX emission spectrum at $P_{\rm ex} = 1$~$\mu$W and $V_{\rm g} = -4$~V. (c) IX emission spectrum at $x = 30$~$\mu$m from (b) with linewidth of 0.57 meV. (d) Spectrally integrated IX emission intensity (laser excitation profile is shown as dashed line), (e) IX emission energy, and (f) IX emission linewidth.}
\end{figure}

We probed the IX transport both by cw (Fig.~2) and time-resolved (Fig.~3) PL imaging. Figure 2a shows a cw $x$-energy emission image. A spectrally wide emission at the origin $x = 0$ corresponds to the $n^+$-GaAs layer, its spatial profile is close to the profile of laser excitation spot with HWHM $\sim 7$~$\mu$m. IXs expand well beyond the laser excitation spot indicating long-range IX transport (Fig.~2a). 

The spatial dependence of IX emission spectrum, spectrally integrated IX emission intensity, IX emission energy and linewidth are shown in Figs.~2b, 2d, 2e, and 2f, respectively. The IX emission shows a ring around the excitation spot (Fig.~2a,b,d). This ring is similar to the inner ring in the emission pattern of IXs in CQW~\cite{Butov2002, Ivanov2006, Hammack2009, Alloing2012}. The enhancement of IX emission intensity with increasing distance from the center originates from IX transport and energy relaxation as follows. IXs cool toward the lattice temperature when they travel away from the laser excitation spot, thus forming a ring of cold IXs. The cooling increases the occupation of the low-energy optically active IX states, producing the IX emission ring. The IX energy reduces with $x$ (Fig.~2e) indicating reducing IX density.

The IX linewidth nonmonotonically varies with $x$ (Fig.~2f). This dependence is in a qualitative agreement with a model~\cite{High2009nl} considering the effects of interaction and disorder on the IX linewidth. Closer to the origin, the IX density is higher and the IX linewidth is larger due to higher interaction-induced broadening~\cite{High2009nl}. Higher IX temperatures at the origin also contribute to the line broadening. Further away from the origin, the IX density is lower and the IX linewidth is larger due to less efficient screening of in-plane QW disorder by IXs and, in turn, higher disorder-induced broadening~\cite{High2009nl}. The narrowest IX emission is observed at some distance away from the excitation spot (around $x \sim 30$~$\mu$m for the data in Fig.~2f). In this area, the IX linewidth lowers to 0.57~meV (Fig. 2c,f). This IX linewidth in WSQW is about 2 times narrower than the IX linewidth in similar experiments in CQW~\cite{Butov2002, Ivanov2006}. The narrow IX linewidth in WSQW indicates a low in-plane disorder, showing the advantages of WSQWs for creating low-disorder IX devices. 

\begin{figure*}[htbp]
\includegraphics[width=\textwidth]{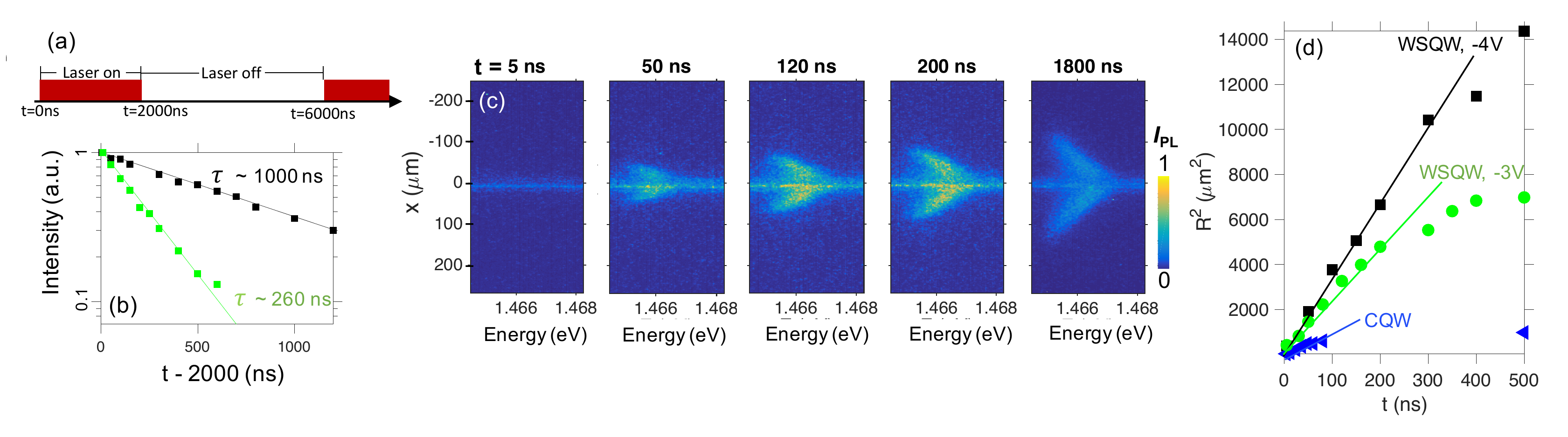}
\caption{(a) Schematic of the rectangular laser excitation pulse profile. Time $t = 0$ corresponds to the onset of the laser pulse. The pulse width $\tau_{\rm width} = 2000$~ns and pulse period $\tau_{\rm pulse} = 6000$~ns. (b) Normalized total IX PL intensity (integrated over $x$ and energy) vs $t$ after laser pulse termination for $V_{\rm g} = - 4$~V (squares) and $- 3$~V (points). The corresponding IX lifetimes are 1000 ns and 260 ns, respectively. (c) $x$-energy IX PL images taken during the laser excitation pulse for several delay times $t$ after the pulse onset. Each image is integrated over a time window of $\delta t = 5$~ns. Laser excitation spot with HWHM $R_0 = 5$~$\mu$m is centered around $x = 0$. (d) The square of HWHM of IX cloud presenting the IX transport radius $R^2$ vs delay time $t$ for $V_{\rm g} = - 4$~V (squares) and $- 3$~V (points). Similar data for IXs in CQW from Ref.~\cite{Dorow2017} are shown by triangles for comparison. Average $P_{\rm ex} = 30$~$\mu$W for WSQW data.}
\end{figure*}

IX transport is also measured by time-resolved PL imaging (Fig.~3). The $x$-energy images in Fig.~3c show the expansion of the IX cloud with time due to IX transport. The time-resolved imaging of the IX cloud expansion enables estimating IX transport characteristics. For a comparison between IX transport in WSQW and IX transport in CQW, which was measured earlier in a similar experiment~\cite{Dorow2017}, the IX transport distance is characterized by the HWHM of the spectrally integrated IX emission, $R$, as in Ref.~\cite{Dorow2017}. Figure 3d shows that at the initial delay times, $R^2$ grows nearly linearly with time $t$. Fitting to the slope by $R^2 = R_0^2 + D^*t$ (solid lines in Fig.~3d) gives an estimate of the IX diffusion $D^* \sim 350$~cm$^2$/s for IXs in WSQW for $V_{\rm g} = - 4$~V and $\sim 300$~cm$^2$/s for IXs in WSQW for $V_{\rm g} = - 3$~V. A similar estimate for IXs in CQW is $\sim 90$~cm$^2$/s~\cite{Dorow2017} (the similar data for IXs in CQW from Ref.~\cite{Dorow2017} are also presented in Fig.~3d for comparison). The IX diffusion coefficient in WSQW is roughly three times higher than in CQW studied earlier. A high diffusion coefficient $D$ corresponds to a high IX mobility $\mu$, which can be estimated using the Einstein relation $\mu = D/(k_{\rm B}T)$, $k_{\rm B}$ is the Boltzmann constant. 

The IX transport experiment is briefly discussed below. Both IX drift and diffusion contribute to the experimentally measured expansion of the IX PL cloud. With increasing distance from the origin, the IX density and, in turn, energy decreases. This creates the IX energy gradient causing the IX drift away from the origin. The IX energy reduction with $x$ is observed in the $x$-energy images (Fig.~3c). Fitting the IX cloud expansion by $R^2 = R_0^2 + D^*t$ probes an effective IX diffusion coefficient $D^* \sim D + \mu n u_0$, which encapsulates both drift and diffusion~\cite{Dorow2017}. $u_0$ originates from the repulsive dipolar interactions causing the IX energy shift $\Delta E = u_0 n$~\cite{Dorow2017}. 

The IX cloud expansion can be approximated by $R^2 \sim R_0^2 + D^{*}t$ at delay times $t$ considerably smaller than the IX lifetimes $\tau$. At $t \sim \tau$ the IX cloud expansion saturates at $R_{\rm sat}^2 \sim R_0^2 + D^{*} \tau$. 

The repulsively interacting IXs screen the in-plane QW disorder, which appears due to the heterostructure imperfactions such as the QW width and composition fluctuations. Since the IX density drops with increasing distance from the origin the IX screening ability and, in turn, $D$ reduces with $x$.

$R$ characterizes the width of the IX PL intensity profile. However the PL intensity profile differs from the IX density profile, in particular due to the IX temperature variation with $x$ leading to the inner ring in IX PL pattern noted above~\cite{Butov2002, Ivanov2006, Hammack2009}. 

Taking these effects into account can improve an estimate of IX diffusion. This is the subject of future work. The similar analysis of IX transport in WSQW and IX transport in CQW probed in the similar experiments (Fig.~3) shows that the IX mobility in WSQW is significantly higher. A high IX mobility in WSQW (Fig.~3) and a narrow IX PL line (Fig.~2c) indicate a low in-plane disorder in WSQW. 

Figure~3b shows that the IXs in WSQW have long and voltage-controllable lifetimes. The IX lifetime 1000~ns at $V_{\rm g} = - 4$~V and 260~ns at $V_{\rm g} = - 3$~V (Fig.~3b) is long enough both for the IX cooling below the temperature of quantum degeneracy~\cite{Butov2001} and for the achievement of long-range IX transport (Figs.~2 and 3) over lengths exceeding the in-plane dimensions of excitonic devices outlined in the introduction.

In this work, GaAs heterostructures are considered. IXs can be realized also in other materials including ZnO, GaN, and van der Waals heterostructures~\cite{Lefebre2004, Morhain2005, Fedichkin2015, Kuznetsova2015, Fedichkin2016, Wang2018}. In these materials, IXs have high binding energies and can be observed up to high temperatures. The possibility to extend a WSQW platform to excitonic devices in different materials is the subject of future work.

In summary, we presented WSQW heterostructures with high IX mobility, spectrally narrow IX emission, voltage-controllable IX energy, and long and voltage-controllable IX lifetime. This set of properties shows that WSQW heterostructures provide an advanced platform both for studying basic properties of IXs in low-disorder environments and for the development of high-mobility excitonic devices.

We thank Michael Fogler for discussions. This work was supported by NSF Grant No.~1640173 and NERC, a subsidiary of SRC, through the SRC-NRI Center for Excitonic Devices. Kinetics measurements were supported by DOE Office of Basic Energy Sciences under award DE-FG02-07ER46449. C.~J.~D. was supported by the NSF Graduate Research Fellowship Program under Grant No.~DGE-1144086. The work at Princeton University was funded by the Gordon and Betty Moore Foundation through the EPiQS initiative Grant GBMF4420, and by the National Science Foundation MRSEC Grant DMR~1420541.

\end{document}